\documentclass[aps,pre,reprint,superscriptaddress]{revtex4-1}

\usepackage{amsmath}	
\usepackage{bm}
\usepackage{ulem}
\usepackage{graphicx}
\usepackage{amssymb}
\usepackage{amsfonts}
\usepackage{latexsym}
\usepackage{enumerate}
\usepackage{alltt}
\usepackage{array}
\usepackage{paralist}
\usepackage{here}
\usepackage{textcomp}

\newcommand{\av}[1]{\langle{#1}\rangle}

\newcommand{\nn}{\nonumber\\}
\newcommand{\del}{\!\!\!}
\newcommand{\px}{\partial_x}
\newcommand{\stem}{\mathrm{stem}}
\newcommand{\TA}{\mathrm{TA}}

\newcommand{\udiv}{u_{\mathrm{div}}}

\newcommand{\udd}{u_{\mathrm{dd}}}
\newcommand{\udm}{u_{\mathrm{dm}}}
\newcommand{\Uder}{U_{\mathrm{der}}}

\begin{document}

 \title{Interplay between epidermal stem cell dynamics and dermal deformations}
 \author{Yasuaki Kobayashi}
 \email{kobayashi.yasuaki@ocha.ac.jp}
 \affiliation{Center for Simulation Sciences, Ochanomizu University, Tokyo 112-8620, Japan }
 \author{Yusuke Yasugahira}
 \affiliation{Graduate School of Science, Hokkaido University, Sapporo 060-0810, Japan}
 \author{Hiroyuki Kitahata}
 \affiliation{Department of Physics, Chiba University, Chiba 263-8502, Japan}
 \author{Mika Watanabe}
 \author{Ken Natsuga}
 \affiliation{Department of Dermatology, Graduate School of Medicine, Hokkaido University, Sapporo 060-8638, Japan}
 \author{Masaharu Nagayama}
 \email{nagayama@es.hokudai.ac.jp}
 \affiliation{Research Institute for Electronic Science, Hokkaido University, Sapporo 060-0812, Japan}

 \date{\today}

 \begin{abstract}
  We introduce a particle-based model of self-replicating cells on a deformable substrate composed of the dermis and the basement membrane and investigate the relationship between dermal deformations and stem cell pattering on it. We show that our model reproduces the formation of dermal papillae, protuberances directing from the dermis to the epidermis, and the preferential stem cell distributions on the tips of the dermal papillae, which the basic buckling mechanism fails to explain. We argue that cell-type-dependent adhesion strength of the cells to the basement membrane is crucial factors of these patterns. 
 \end{abstract}
 \maketitle
 \section{Introduction}
 The morphology of growing tissues is influenced by mechanical forces due to cell division, migration, and apoptosis \cite{Vincent,hannezo2,shraiman,mammoto,ranft}, where the buckling instability is regarded as an important pattern-forming mechanism \cite{drasdo,hannezo_pnas}. Buckling is quite often observed when a growing tissue layer interacts with an elastic substrate or two growing layers with different growth rates interact with each other \cite{basan2,Tallinen, nelson}. Many biological systems have been modeled from such a viewpoint, examples including airway epithelium \cite{Varner}, intestine \cite{hannezo, Ben_Amar, shyer}, colonic crypt \cite{Dunn}, and tumor \cite{Ciarletta}. 

\begin{figure}[tb]
  \centering
  \includegraphics[width=\linewidth]{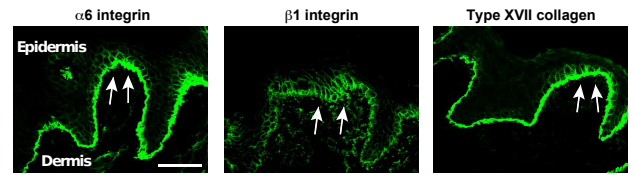}
  \caption{Expression of stem cell markers in normal human epidermis. Labeling of epidermal stem cell markers ($\alpha$6 integrin, $\beta$1 integrin and type XVII collagen) is pronounced at the tip of dermal papillae (indicated by arrows). Scale bar: 50 $\mu$m.}\label{fig:exp}
 \end{figure}

 Skin provides another example of such systems, where a sheet of proliferating cells (the basal layer) is attached via the basement membrane to a soft elastic substrate (the dermis). Its morphology has two distinct features: First, the interface between the basal layer and the dermis has many protuberances directing outward (namely from dermis to epidermis), which are called \textit{dermal papillae}. Second, several experiments suggest that stem cells tend to be found on the tips of dermal papillae, while the transit amplifying cells occupy the rest of the dermal surface \cite{jones, jensen2}, which we have experimentally confirmed, as shown in Fig.~\ref{fig:exp}. Although buckling mechanism is considered to be responsible for the formation of such dermal undulations \cite{Kuwazuru}, it alone does not explain why dermal papillae direct outward rather than inward, nor does it tell us why stem cells prefer the tips of dermal papillae.

 Skin is an important organ that provides us with barriers such as protecting from damages and preventing dehydration \cite{Elias, natsuga}, and its internal structure affects barrier functions: for example, diseases such as psoriasis are accompanied by altered structure of the epidermal-dermal interface, where the germinative cell population increases and their activity is enhanced \cite{weinstein, Iizuka95}. Although mathematical models have been proposed to understand skin barrier functions \cite{kobayashi1,kobayashi2,Denda_koba,kobayashi3,kobayashi4,Suetterlin,suetterlin2,walker,Schaller}, the morphology of the epidermal-dermal interface and its relationship to stem cell patterning has been largely ignored, except for a speculation on the interplay between stem cell activities and the dermal structure \cite{Iizuka96}. 

 In this work we introduce a mathematical model of cell division dynamics in the basal layer on a substrate composed of the basement membrane and the dermis, taking into account substrate deformability and cell type-dependent adhesion to the substrate. We employ particle-based modeling that has been used to study growing tissues \cite{basan3, drasdo} and the whole epidermal simulation \cite{Suetterlin,kobayashi1,kobayashi4}. By numerical experiments we show that our model can reproduce dermal papillae formation and preferential stem cell patterning on the tips of the dermal papillae. We argue that the outward deformation of the dermis is caused by the interplay of cell-membrane adhesion and membrane elasticity, and that the stem cell distribution is determined by the difference in adhesion strength between different types of cells.

 \section{The model}\label{sec:model}
 \begin{figure}[tb]
  \begin{center}
   \includegraphics[width=0.9\linewidth]{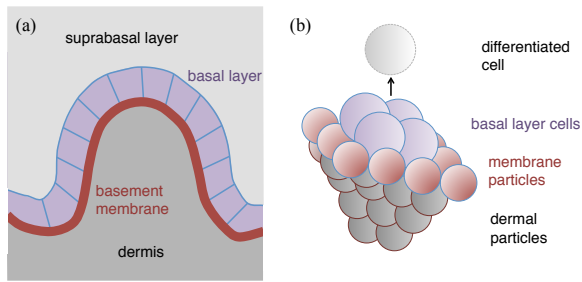}
   \caption{(a) Inner structure of the skin. The basal layer is made of undifferentiated cells. The suprabasal layer is made of differentiated cells (individual cells not depicted). (b) Particle-based modeling of the skin. Dermal particles and membrane particles correspond to the dermis and the basement membrane, respectively. The basal layer cells repeat division on the basement membrane. Differentiated cells, which would form the suprabasal layer, are removed from the system.  }\label{fig:schematic}
  \end{center}
 \end{figure}
 The inner structure of the skin is schematically shown in Fig.~\ref{fig:schematic}(a). The dermis and the basement membrane form an elastic substrate, and the basal layer cells are attached to the basement membrane. These cells repeat division on this substrate and, when differentiated, they migrate and join the suprabasal layer. 

 To investigate these cell dynamics and their possible effects on the substrate shape, we employ the following particle-based model, a schematic of which is shown in Fig.~\ref{fig:schematic}(b). Both the dermis and the basement membrane are approximated by different kinds of spheres, which we call \textit{dermal particles} and \textit{membrane particles}, respectively. Dermal particles, with radius $R_d$, are considered as independent particles, whereas membrane particles, with radius $R_m$, are connected by links to form a triangular lattice network. On this substrate we consider two kinds of cells, stem cells and transit amplifying (TA) cells, both represented as spheres with radius $R_c$. We assume the following difference between the two: A stem cell divides into a stem cell and a TA cell, while a TA cell divides into two TA cells; stem cells can divide an infinite number of times, while TA cells can divide only a finite number of times; Stem cells are strongly bound to the basement membrane, while TA cells are weakly bound so that they are able to detach from the basement membrane when they undergo differentiation (conditions for differentiation is described below). For simplicity, we disregard the suprabasal layer in this model: differentiated cells are removed from the system.
  
  Equations of motion for dermal particles, membrane particles, and the two kinds of cells are obtained by specifying interactions between them, together with the rules of cell division, which are explained below.
  
  \begin{figure}[tb]
   \begin{center}
    \includegraphics[width=0.8\linewidth]{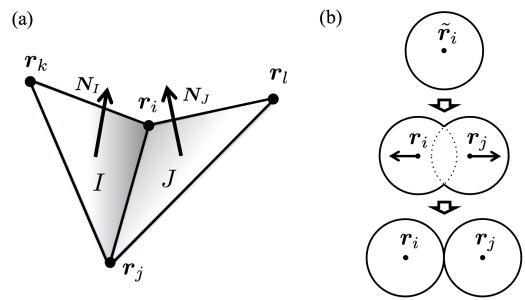}
    \caption{(a) Two adjacent triangular regions of the membrane network sharing a link $(i, j)$, with unit normals $\bm{N}_{ij}^{(1)}$ and $\bm{N}_{ij}^{(2)}$ defined by four network-connected membrane particles $i$, $j$, $k$, and $l$. (b) Cell division process. Cell $i$, initially at $\tilde{\bm{r}}_i$, is dividing into two cells $i$ and $j$. Division is completed when the natural length of the spring between the cells, which grows in time, exceeds $2R_c$. }\label{fig:model}
   \end{center}
  \end{figure}

  \subsection{The dermis}
  A dermal particle can interact with (i) another dermal particle, (ii) a membrane particle, and (iii) the boundary wall, which is the lower boundary of the system. Let $d$ be the distance between the two interacting entities. In the case (i), the interaction is strongly repulsive when $d < D\delta_1$, weakly repulsive when $D\delta_1 \le d < D$, weakly adhesive when $D \le d < \Lambda D$, and zero when $\Lambda D \le d$, where $D=2R_d$ is the contact distance of the two particles and the parameters ${\delta_1}$ and $\Lambda$ control the compressibility of the dermis and the cutoff range of the long-range interaction, respectively. This interaction is determined in such a way that the dermal particles as a whole exhibit plasticity and elasticity. In the case (ii), the interaction is the same as in the case (i), except that $D=R_d+R_m$ and ${\delta_1}=1$: there is no intermediate weak-repulsion range for the dermis-membrane interaction. In the case (iii), the interaction is repulsive only when the particle overlaps with the boundary and otherwise zero. 

  The total energy of dermal particles associated with these interactions is represented by $\Uder$, which is defined in Appendix \ref{sec:dermis}.
  
  \subsection{Membrane deformation}
  Elasticity of the basement membrane is introduced by assigning the stretching energy and the bending energy to individual links of the triangular membrane network. For a given link, the stretching energy is considered between the two particles with distance $d$ connected by this link, which is harmonic in the range $2R_m{\delta_2} < d < 2R_m$ and steeply increases outside of this range, so that too much stretching and compressing are prevented. 

  The bending energy for a given link is considered between the two triangular regions sharing this link [see Fig.~\ref{fig:model}(a)]. Here we assume that, as a function of the angle $\theta$ between the unit normals of the two triangular regions, the bending energy is proportional to $\theta^4$ for small $\theta$, which is softer than $\theta^2$ as is usually used for elastic plates: This allows the substrate to exhibit local deformations, while preventing folding of the basement membrane for large deformations. 

  In addition, short-range repulsive interactions are required between membrane particles that are not network-connected so that their overlapping is avoided when the membrane is largely deformed. The total energy associated with these interactions is represented by $U_{\mathrm{mm}}$, which is defined in Appendix \ref{sec:memb}. 
  
  \subsection{Cell-basement membrane adhesion and cell differentiation}
  Both stem cells and TA cells repulsively interact with membrane particles when they overlap. In addition, each cell adhesively interacts only with its nearest membrane particle. Note that, for a given cell, its nearest partner can change by the movement of the surrounding cells and membrane particles, so that cells can move along the surface of the basement membrane while keeping adhesiveness by changing their adhesive interaction partners. We assign different adhesive forces to stem cells and TA cells so that only TA cells can be detached from the basement membrane, whereas stem cells remain in the basal layer.  
  
  We identify differentiation of TA cells with mechanical detachment from the basement membrane: When a TA cell and its nearest membrane particle with distance $d$ satisfies $d > \lambda(R_c+R_m)$, this cell is regarded as differentiated and removed from the system. Here $\lambda$ is a parameter determining the differentiation threshold. We assume that non-proliferative TA cells, namely those which have lost the ability of cell division, lose adhesiveness to the basement membrane so that they are easily pushed away from the basal layer. Such a relationship between proliferation and adhesiveness has been experimentally observed \cite{jones2,watanabe}. The total energy of the cell-basement membrane interactions is represented by $U_{\mathrm{mc}}$, which is defined in Appendix \ref{sec:adhesion}. 

  \subsection{Cell division and cell-cell interactions}
  Cell division is modeled as a process of two initially overlapping cells being gradually separated [see Fig.~\ref{fig:model}(b)], which is different from our previous work \cite{kobayashi1} and similar to Ref.~\cite{drasdo}. In this description, when cell $i$ starts division, a new cell $j$ is immediately introduced into the system, where the two cells are almost completely overlapped, with their initial positions given by $\bm{r}_j=\tilde{\bm{r}}_i+ \epsilon R_c\bm{t}$ and $\bm{r}_i=\tilde{\bm{r}_i} - \epsilon R_c\bm{t}$, where $\tilde{\bm{r}_i}$ is the position of $i$ just before the division, $\epsilon$ a small parameter, and $\bm{t}$ a unit vector that is randomly oriented within a plane perpendicular to the axis between $\tilde{\bm{r}}_i$ and its nearest membrane particle, thus approximating a tangent vector of the basement membrane at $\tilde{\bm{r}}_i$. Let us call the two cells committing a division process a \textit{dividing pair}. A dividing pair is connected with a spring whose natural length grows in time from the initial value $2\epsilon R_c$. The division is completed when the natural length reaches $2\epsilon R_c$; At this point the spring interaction of the dividing pair is dissolved. 

  Interaction of two cells (with distance $d$) that do not form a dividing pair is repulsive when $d<2R_c$ and adhesive when $2R_c\le d < 2\Lambda R_c$. Note that each cell in a dividing pair behaves as an independent cell when it interacts with other cells or membrane particles. The total energy associated with cell division and cell-cell interactions is given by $U_{\mathrm{cc}}$ defined in Appendix \ref{sec:cell-cell}. Modeling of cell cycle is the same as in the previous work \cite{kobayashi1} and is explained in Appendix \ref{sec:cellcycle}. 
  
  \subsection{The equations of motion}
  The equations of motion are written down by using the above-introduced interaction energies as
  \begin{align}
   \frac{d\bm{r}_i}{dt}
   &= -\frac{\partial}{\partial \bm{r}_i}U_{\mathrm{der}} , \quad i\in \Omega_d, \\
   \frac{d\bm{r}_i}{dt}
   &= -\frac{\partial}{\partial \bm{r}_i}\left(U_{\mathrm{der}}+U_{\mathrm{mm}}+U_{\mathrm{mc}} \right), \quad i\in \Omega_m,\\
   \frac{d\bm{r}_i}{dt} &= -\frac{\partial}{\partial \bm{r}_i}\left(U_{\mathrm{mc}}+U_{\mathrm{cc}} \right), \quad i\in \Omega_c,\label{eq:stem}
  \end{align}
  where $\Omega_d$, $\Omega_m$, and $\Omega_c$ are the sets of dermal particles, membrane particles, and cells, respectively, and $\bm{r}_i=(x_i,y_i,z_i)$ is the position of $i$ in each set. Note that the number of TA cells in the basal layer changes in time due to cell division and differentiation, whereas the number of stem cells remains unchanged. 

  The model is considered in a three-dimensional region defined as $0\le x < L_x$, $0\le y < L_y$, and $0\le z < L_z$, with periodic boundaries in the $x$ and $y$ directions. The boundary wall supporting the whole system is placed at $z=0$. The system size is chosen as $L_x=L_y=50$ and $L_z=100$. For the basement membrane, $200\times 230$ membrane particles are arranged to form a triangular network on the plane $z=16R_c$. They are initially partially overlapped, with distance approximately equal to $2R_m{\delta_2}$. Dermal particles are placed between the basement membrane and the boundary wall by random packing. The radii are chosen as $R_c=1.4$, $R_m=1.0$, and $R_d=1.4$, so that neither cells or dermal particles can pass through the membrane network. The other parameters are set to ${\delta_1}=0.8$, ${\delta_2}=0.25$, $\Lambda=1.4$, and $\lambda=1.4$. 

  \section{Simulation results}\label{sec:results}
  \subsection{Formation of dermal papillae}
\begin{figure*}[tb]
   \begin{center}
    \includegraphics[width=.7\linewidth]{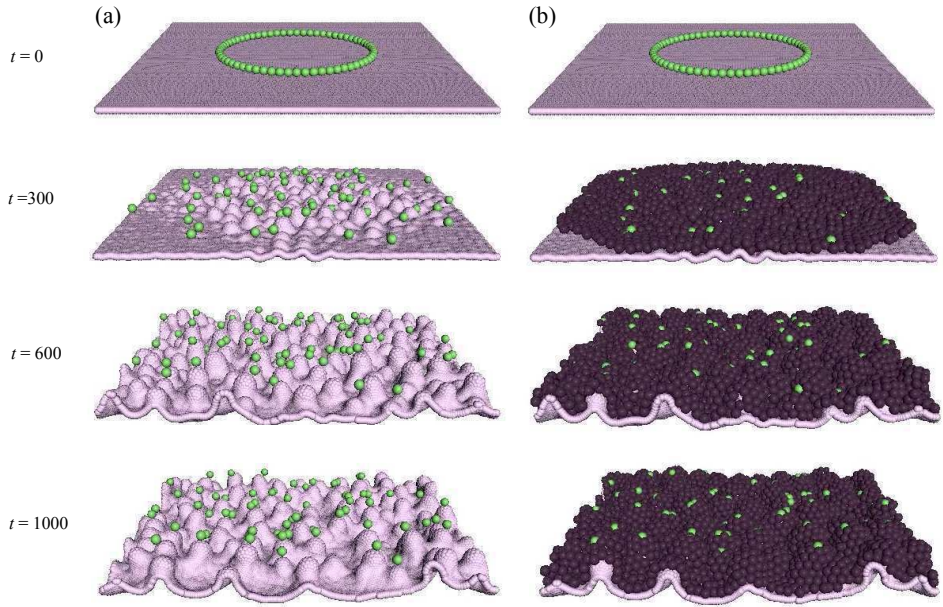}
    \caption{Simulation snapshots of dermal papillae formation. Stem cells are initially placed in a ring arrangement. (a) and (b) show the same simulation results, where (a) only stem cells (light green) are visualized, or (b) both stem cells and TA cells (dark purple) are visualized. In both cases dermal particles are not visualized. }\label{fig:normal_snap}
   \end{center}
  \end{figure*}

  First we perform numerical simulations of our model, starting with an initially flat basement membrane on which 64 stem cells are placed in a certain arrangement. Figure \ref{fig:normal_snap} shows a simulation result in the case of a ring arrangement of the stem cells. As time evolves, TA cells are continually produced first by stem cells and then by themselves, spreading over the surface of the basement membrane and soon entirely cover it to form the basal layer, while the basement membrane starts to deform in the regions where it is covered with TA cells. Finally many outward (upward in this case) protuberances are produced, which can be regarded as dermal papillae. During this process, stem cells also spread over the basement membrane. Noticeably, stem cells tend to be located on the tips of the dermal papillae. These observations show that our model reproduces basic features of the membrane shape and stem cell distributions observed in real skin. 
  
  We introduce the following statistical quantities to characterize these spatial patterns.
  First, the mean and Gaussian curvatures of the basement membrane at the position of cell $i$, denoted by $\kappa_H(\bm{r}_i)$ and $\kappa_G(\bm{r}_i)$, respectively, are evaluated from the membrane particle nearest to $i$ and its link-connected membrane particles, using standard formulas of discrete differential geometry \cite{meyer}. Then, to quantify the difference of these curvatures between stem cells and TA cells, we define the normalized curvature differences as follows:
  \begin{align}
   \chi_H &= \bigg(\frac{1}{N_{\mathrm{stem}}} \! \sum_{i\in\Omega_{\mathrm{stem}}} \del \kappa_H(\bm{r}_i)
   -\frac{1}{N_c}\sum_{i\in\Omega_c}\kappa_H(\bm{r}_i)\bigg)R_c,\label{eq:chih}\\
   \chi_G &=\bigg(\frac{1}{N_{\mathrm{stem}}} \! \sum_{i\in\Omega_{\mathrm{stem}}} \del \kappa_G(\bm{r}_i)
   -\frac{1}{N_c}\sum_{i\in\Omega_c}\kappa_G(\bm{r}_i)\bigg)R_c,\label{eq:chig}
  \end{align}
  where $\Omega_{\mathrm{stem}}$ is the set of stem cells, $N_{\mathrm{stem}}$ is the number of stem cells, and $N_c$ is the number of all basal layer cells (stem cells and TA cells). These quantities are the average membrane curvatures at the position of stem cells relative to all basal layer cells, non-dimensionalized by the cell radius $R_c$. 

  In addition, relative height of the stem cells (non-dimensionalized by $R_c$) is defined as 
  \begin{align}
   \zeta=\frac{1}{R_c}\left(
   \frac{1}{N_{\mathrm{stem}}} \! \sum_{i\in\Omega_{\mathrm{stem}}} \del z_i
   -\frac{1}{N_{c}} \sum_{i\in\Omega_{c}} z_i
   \right), \label{eq:height}
  \end{align}
  and the spatial inhomogeneity of the stem cell distribution in the $x$ and $y$ directions is defined as
  \begin{align}
   \sigma_x= \left| \frac{1}{N_\mathrm{stem}} \! \sum_{i \in \Omega_{\mathrm{stem }}} \del 
   e^{\frac{2\pi i x_i}{L_x}} 
   \right|, \quad
   \sigma_y= \left| \frac{1}{N_\mathrm{stem}} \! \sum_{i \in \Omega_{\mathrm{stem }}} \del 
   e^{\frac{2\pi i y_i}{L_x}} 
   \right|, \label{eq:sigma}
  \end{align}
  from which we measure $\sigma=\sqrt{\sigma_x^2+\sigma_y^2}$. 

  \begin{figure}[tb]
   \begin{center}
    \includegraphics[width=\linewidth]{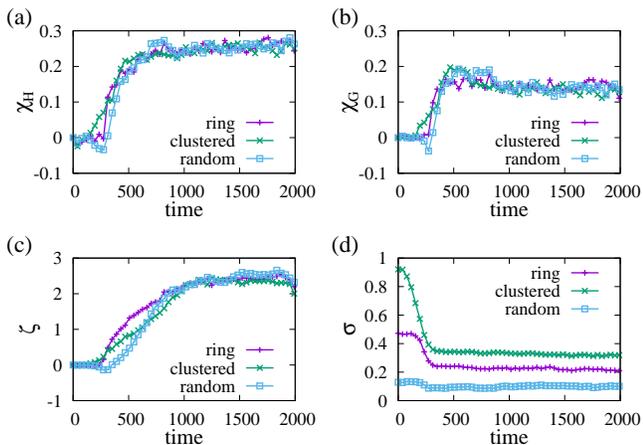}
    \caption{Time courses of statistical quantities characterizing membrane deformation and stem cell distributions: (a) Normalized mean curvature difference $\chi_H$ [\eqref{eq:chih}], (b) Normalized Gaussian curvature difference $\chi_G$ [\eqref{eq:chig}], (c) Normalized height difference $\zeta$ [\eqref{eq:height}], and (d) Spatial inhomogeneity $\sigma=\sqrt{\sigma_x^2+\sigma_y^2}$ [\eqref{eq:sigma}]. Results are shown for three initial stem cell arrangements: ring-shaped (corresponding to Fig.~\ref{fig:normal_snap}), clustered, or random. }\label{fig:normal_stats}
   \end{center}
  \end{figure}

  Figure \ref{fig:normal_stats} shows time evolution of these quantities for three different initial stem cell arrangements: the ring arrangement as shown in Fig.~\ref{fig:normal_snap}; a clustered arrangement, where all stem cells are gathered at the center; and a random arrangement. We find that the normalized curvature differences $\chi_H$ and $\chi_G$ and the normalized height difference $\zeta$ increase from zero and reach stationary values, which are significantly larger than zero, which means that stem cells tend to be located on the tips of the dermal papillae. Furthermore, $\chi_H$ and $\chi_G$ are faster than $\zeta$ to converge to the stationary states, indicating that small papillae are created at the position of stem cells first and then their sizes increase, not that the full-grown papillae are created first and then stem cells move toward their tips. Increase of $\chi_H$, $\chi_G$, and $\zeta$ starts at around $t\sim 300$, at which the entire surface of the basement membrane is covered with TA cells, as shown in Fig.~\ref{fig:normal_snap}(b). This implies that dermal papillae formation is caused by the pressure created by continuous division of cells, which is weak when there is enough room for spreading of TA cells.

  The quantities $\chi_{\mathrm{H}}$, $\chi_{\mathrm{G}}$, and $\zeta$ show no noticeable dependence on initial stem cell arrangements. In contrast, the spatial inhomogeneity measure $\sigma$ shows this dependence, as shown in Fig.~\ref{fig:normal_stats}(d). Here the clustered arrangement, which is initially highly inhomogeneous, results in large $\sigma$ values, whereas the random arrangement, which is initially already homogeneous, keeps small $\sigma$ values, although in all three cases $\sigma$ decreases in time and reaches stationary values at around $t\sim 300$. This decrease in the early stage is because spreading of TA cells causes stem cells to separate from each
  other. In the later state, stem cells settle in the tips of the dermal papillae and resist further separation, resulting in the residual values of $\sigma$. 
  
  \subsection{Numerical experiments by introducing non-dividing cells}\label{sec:nd}
  The observed preference of stem cells to occupy the tips of the dermal papillae must be attributed to the difference between stem cells and TA cells, which in this model is either the way of adhesion to the basement membrane (strongly bound or detachable) or the ability of cell division (infinite or finite number of times).  To explore this we introduce \textit{non-dividing} (ND) cells, which have the same type of adhesion to the membrane as stem cells (strongly bound to and never detach from the basement membrane) but do not replicate at all; their dynamics are described by \eqref{eq:stem}. 
  
  \begin{figure*}[tb]
   \begin{center}
    \includegraphics[width=\linewidth]{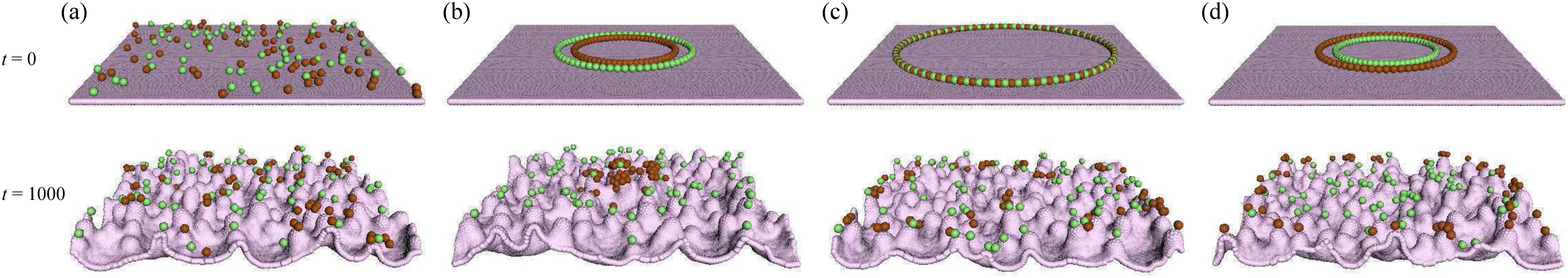}
    \caption{Numerical experiments with stem cells (light green) and non-dividing (ND) cells (dark red) in four different initial arrangements, as shown in top panels. Simulation results at $t=1000$ are shown in bottom. }
    \label{fig:nondivexp}
   \end{center}
  \end{figure*}
  We place the same number of stem cells and ND cells (64 for each) on the basement membrane in four different initial arrangements, as shown in the top panels of Fig.~\ref{fig:nondivexp}. Results of numerical simulations at $t=1000$ are shown in the bottom panels of Fig.~\ref{fig:nondivexp}. We find that both stem cells and ND cells prefer the tips of dermal papillae. Compared to stem cells, ND cells tend to share the same papilla and sometimes form one large cluster as shown in Fig.~\ref{fig:nondivexp}(b). 
  
\begin{figure*}[tb]
   \begin{center}
    \includegraphics[width=\linewidth]{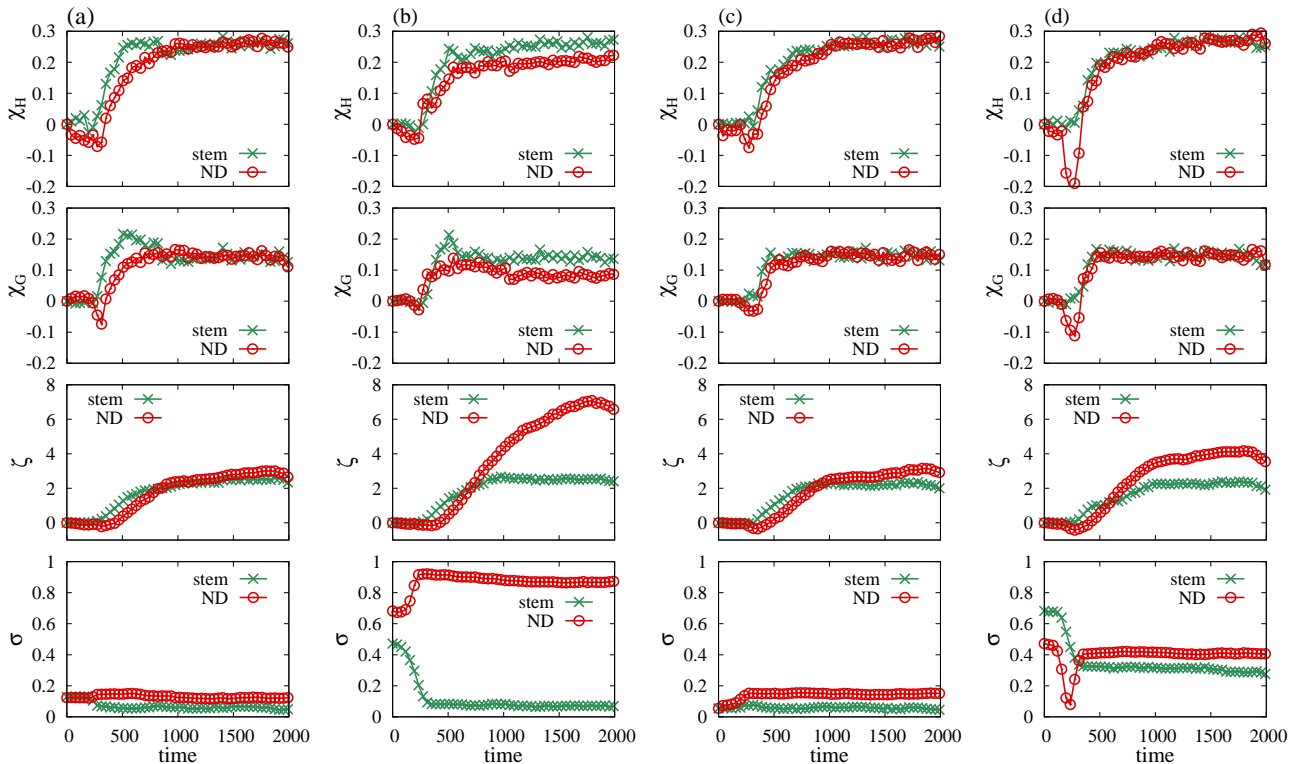}
    \caption{Time courses of normalized mean curvature difference $\chi_{\mathrm{H}}$, normalized Gaussian curvature difference $\chi_{\mathrm{G}}$, normalized height difference $\zeta$, and spatial inhomogeneity $\sigma$, for stem cells and non-dividing (ND) cells. The labels (a), (b), (c), and (d) correspond to the four different initial arrangements of stem cells and ND cells with the same labels in Fig.~\ref{fig:nondivexp}. 
    }\label{fig:nd_stats}
   \end{center}
  \end{figure*}

  In Fig. \ref{fig:nd_stats} we plot $\chi_H$, $\chi_G$, $\zeta$, and $\sigma$ for stem cells and ND cells: For the latter, $\Omega{\mathrm{stem}}$ and $N_{\mathrm{stem}}$ in these quantities are replaced by the set of ND cells and the total number of them, respectively. We find that significantly large positive values of $\chi_H$ and $\chi_G$ are observed for both stem cells and non-diving cells, indicating that these cells are located on the tips of dermal papillae. There is no noticeable difference in $\chi_H$ and $\chi_G$ between stem cells and ND cells, nor is there any difference between different initial arrangements. 
  
  Difference between stem cells and ND cells are captured in $\zeta$ and $\sigma$. The height difference $\zeta$ in ND cells is slightly smaller than that in stem cells in the early stage for all four initial arrangements, presumably because there are fewer cell division events in the vicinity of ND cells so that they feel smaller forces that can cause the membrane to locally deform. In the later stage, $\zeta$ in ND cells becomes larger than that in stem cells, especially in the case of Figs.~\ref{fig:nd_stats}(b) and (d). This tendency correlates with spatial inhomogeneity: The inhomogeneity measure $\sigma$ in ND cells is significantly higher than that in stem cells. In particular, Figs.~\ref{fig:nd_stats}(b) and \ref{fig:nd_stats}(d) show that ND cells have a stronger tendency to preserve initial inhomogeneity than stem cells, which reflects the observed clustering of ND cells in Figs.~\ref{fig:nondivexp}(b) and \ref{fig:nondivexp}(d) (note the periodic boundaries). The correlation between $\zeta$ and $\sigma$, together with the slower increase of the former, indicates that clustering of ND cells enhances the formation of large outward protuberances. 

  \section{Discussions}\label{sec:discuss}
  The results of the previous section indicate that the strength of adhesion to the basement membrane is crucial for the observed patterns. Now the following argument explains why protuberances of the basement membrane direct outwards. Let us consider a small membrane segment on which cells are attached. When a cell division occurs and the surface of the segment becomes crowded, either a cell has to detach from the segment or the segment has to deform so that it can accommodate all attached cells, depending on the energy cost for detachment and deformation. If deformation is energetically more preferable than detachment, the resulting shape of the segment will be convex so as to minimize the stretching energy. Thus from a local point of view every place has a tendency to create outward protuberances, resulting in dermal papillae. 

  This energetic argument also explains why stem cells prefer the tips of the dermal papillae: 
  Cells feel the stronger pressure to push them out of the basal layer in the concave regions such as the bottom of the basement membrane than in the convex regions such as the top, which means that the larger cost is required to remain in the basal layer in the former case. In such situations the total cost can be minimized when strongly bound cells such as stem cells occupy the top and TA cells fill the remaining space. 
  
  A coarse-grained continuum description helps us understand this situation. For simplicity, let us consider a one-dimensional basement membrane, whose deviation (assumed to be small) from the flat surface is denoted by $h(x,t)$, on which cells with different adhesion strength repeat division at a uniform rate. For a membrane segment in the region $[x,x+{\Delta} x]$ with the length $\Delta s$ and the curvature $\kappa$, the adhesion energy between the cells and the segment is proportional to the length of the cell layer, which is estimated as $\Delta s(\kappa^{-1}+R_c)/\kappa^{-1}$, which reflects the fact that a convex shape can accommodate more cells than a concave shape of the same segment. Substituting ${\Delta} s=[1+\frac{1}{2}(\px h)^2]{\Delta} x$ and $\kappa=-\px^2 h$ at the lowest order of $h$ and taking the limit ${\Delta} x\to 0$, we obtain the energy density associated with the adhesion, denoted by $\epsilon_a$:
  \begin{align}
   \epsilon_a &= -K_a(x)\bigg[1+\frac{1}{2}\bigg(\frac{\partial h}{\partial x}\bigg)^2\bigg]
   +R_c K_a(x)\frac{\partial^2 h}{\partial x^2},
  \end{align}
  where $K_a(x)$ represents the difference in adhesion strength depending on cell types. Then the free energy functional can be written as $F=\int (\epsilon_a + \epsilon_r) dx$, where $\epsilon_r$ represents the remaining interactions including elasticity of the substrate. By taking variation with regard to $h$, we arrive at
  \begin{align}
   \tau\frac{\partial h}{\partial t} = -\frac{\partial }{\partial x}\bigg(K_a(x)\frac{\partial h}{\partial x}\bigg)
   - R_c \frac{d^2K_a}{dx^2} + \mathcal{R}, \label{eq:continuum}
  \end{align}
  where $\tau$ is a constant and $\mathcal{R}$ comes from $\epsilon_r$. Thus the cell-substrate adhesion gives rise to the first two terms, both contributing to destabilizing the uniform state $h=0$, which is balanced by the substrate elasticity included in $\mathcal{R}$. 

  The first term is regarded as the inverse diffusion. This term is symmetric with respect to upward or downward deviations of $h$ and its effect corresponds to buckling-like mechanism proposed in the previous works. In contrast, the second term is not symmetric in this sense and is responsible for both the formation of outward protuberances and the tip preference of stem cells: 
  For example, when a strongly bound cell such as a stem cell is surrounded by weakly bound cells, in the continuum description $K_a(x)$ has a unimodal shape with a positive peak at the position of the strongly bound cell and $K''_a<0$ holds near this peak. According to \eqref{eq:continuum}, this implies upward deviation of the substrate near a strongly bound cell. The same mechanism also works among TA cells when a cell division locally increases the adhesion cost at the position of the dividing TA cell, thereby creating the same unimodal shape of $K_a(x)$ there. This explains outward protuberances in the places where stem cells are absent, as observed in Fig.~\ref{fig:normal_snap}(a). Note that this second term appears only when we explicitly take into account the size of the cells, which vanishes when $R_c=0$. 

  It has been postulated \cite{Iizuka96} that the pressure created by reproducing cells is responsible for spatial patterning of stem cells. Surely this effect accounts for the spatial distancing of stem cells, clustering of ND cells by failing to create proliferating TA cells around them, and the accompanying deformation of the basement membrane between stem cells. However, it explains neither the growing direction of the dermal papillae nor preferential stem cell locations on their tips, which can be explained only when we take into account the differential adhesion strength between cells. 

  \section{Concluding remarks}\label{sec:conc}
  We have shown that our model can reproduce basic features of stem cell distribution and the dermal structure, especially the formation of dermal papillae and preferential stem cell locations on their tips. We conclude that both the outward growth of dermal papillae and the tip-preference of stem cells are the result of differential cell adhesion to the basement membrane. It has been recently reported that this tip-preference of stem cells could be experimentally reproduced on an elastic sheets with artificial papillae \cite{Viswanathan}, which showed that no external signal from dermis was required for stem cell patterning, thus corroborating our adhesion-based patterning mechanism. Such a generic mechanism is expected to work in other pattern-forming systems. 

  Our results may shed light on photoaging, a physiological senescence process induced by ultraviolet exposure: Ultraviolet destroys elastic fibers in the dermis, leading to dermal stiffness. Photoaging causes changes in the dermal structures, such as flattening of the basement membrane and thinning of the epidermis \cite{Fenske}, accompanied by the decrease in the number and the activity of stem cells \cite{Oh}. This phenomenon is in accordance with our model, which predicts that when elasticity of the basement membrane is lost or the dermis is hardened by photoaging, undulations are suppressed and as a result the thickness of the whole epidermis is reduced because of the diminishing surface area of the basement membrane accommodating fewer TA cells. Numerical experiments of the whole epidermis with varying membrane stiffness would be possible by incorporating the present model into our previous epidermal model \cite{kobayashi1, kobayashi2}. 

  \section*{Acknowledgment}
  This work was supported by JST CREST Grant Number JPMJCR15D2, Japan, and the Cooperative Research Program of ``Network Joint Research Center for Materials and Devices'' (No. 20173006).
  
  \appendix
  \section{Experimental details}
  Specimens from normal human skin samples were frozen on dry ice in an OCT compound. 5 $\mu$m frozen sections were stained without fixation by the following primary antibodies overnight at 4\textcelsius: anti-$\alpha$6 integrin (BD Biosciences, San Diego, CA, USA; GoH3), anti-$\beta$1 integrin (Chemicon International, Billerica, MA, USA) and anti-anti-human type XVII collagen (TS39-3) \cite{ujiie}. After washing in phosphate-buffered saline, the sections were incubated with secondary antibodies conjugated with Alexa488 for 1 hr at room temperature. All the stained sections were observed using a confocal microscope (Olympus Fluoview FV1000; Olympus, Tokyo, Japan). The institutional review board of Hokkaido University Graduate School of Medicine approved the procedures. Participants or their legal guardians gave their written informed consent.

  \section{Interaction potentials between the particles} \label{sec:equations}
  \subsection{The dermis}\label{sec:dermis}
\begin{figure}[tb]
   \begin{center}
    \includegraphics[width=\linewidth]{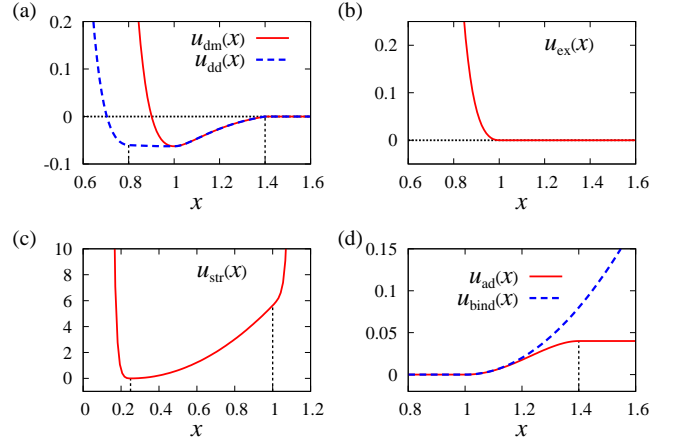}
    \caption{Interaction functions between particles and cells: (a) $u_{\mathrm{dm}}$ (red solid curve) and $u_{\mathrm{dd}}$ (blue dashed curve). The two vertical dotted lines indicate $x=\delta_1$ (left) and $x=\Lambda$ (right) . (b) $u_{\mathrm{ex}}$ (red solid curve). (c) $u_{\mathrm{str}}$ (red solid curve). The region between the two vertical dotted lines indicates the harmonic potential region. (d) $u_{\mathrm{ad}}$ for TA cells (red solid curve) and $u_{\mathrm{ind}}$ for stem cells (blue dashed curve). The vertical line at $x=\lambda$ correspond to the interaction threshold for TA cells, above which the cells are detached from the basement membrane.}\label{fig:potential}
   \end{center}
  \end{figure}

  The total interaction energy of dermal particles with other dermal particles, membrane particles, and the boundary wall, is the following:
  \begin{align}
   U_{\mathrm{der}}=& \del\del \sum_{i\in\Omega_m, j\in\Omega_d}\del\del K_{\mathrm{dm}}
   u_{\mathrm{dm}}\left(\frac{|\bm{r}_i-\bm{r}_j|}{R_m+R_d}\right) \nonumber\\
   & +\del\del \sum_{i, j\in\Omega_d, i<j} \del\del K_{\mathrm{dd}} 
   u_{\mathrm{dd}} \left(\frac{|\bm{r}_i-\bm{r}_j|}{2R_d}\right) \nonumber\\ &+ \sum_{i\in\Omega_d}K_{\mathrm{ex}}
   u_{\mathrm{ex}}\left(\frac{|\bm{r}_i-\hat{\bm{r}}_i|}{2R_d}\right), \label{eq:uder}
  \end{align}
  where $\Omega_m$ and $\Omega_d$ are the sets of membrane particles and dermal particles, respectively, and $\hat{\bm{r}}_i$ is a mirror image of $\bm{r}_i$ regarding the boundary wall at $z=0$. The interaction functions $u_{\mathrm{dm}}$, $u_{\mathrm{dd}}$, and $u_{\mathrm{ex}}$ are defined as
  \begin{gather}
   u_{\mathrm{dm}}(x) = 
   \begin{cases}
    \phi(x)-\phi(\Lambda), & x<\Lambda, \\
    0, & \Lambda \le x,
   \end{cases} 
   \label{eq:dm} \\
   u_{\mathrm{dd}}(x) =
   \begin{cases}
    \frac{\alpha_1}{2}(x-1)^2 + \phi(x+1-{\delta_1}) -\phi(\Lambda), & x < \delta_1, \\
    \frac{\alpha_1}{2}(x-1)^2 +\phi(1)-\phi(\Lambda), & \delta_1 \le x < 1,\\
    \phi(x) -\phi(\Lambda), & 1 \le x < \Lambda, \\
    0, & \Lambda \le x,
   \end{cases} 
   \label{eq:udd} \\
   u_{\mathrm{ex}}(x) =  
   \begin{cases}
    \phi(x) -\phi(1), & x<1, \\
    0, & x \ge 1,
   \end{cases} 
   \label{eq:ex}
  \end{gather}
  where
  \begin{align}
   \phi(x)=\frac{1}{12}x^{-12}-\frac{1}{6}x^{-6}. \label{eq:LJ}
  \end{align}
  Note that $\udm=\udd$ for $x\ge 1$: See Fig.~\ref{fig:potential}(a) and (b). We choose $K_{\mathrm{dd}}=0.04$, $K_{\mathrm{dm}}=0.04$, ${\delta_1}=0.8$, and $\alpha_1=0.125$.

  \subsection{Membrane deformation} \label{sec:memb}
  The total energy of the membrane deformation is given by 
  \begin{align}
   U_{\mathrm{mm}} = &\sum_{(i,j)\in E_m}\del \bigg[ K_{\mathrm{str}} 
   u_{\mathrm{str}} \bigg(\frac{|\bm{r}_i-\bm{r}_j|}{2R_m}\bigg) \nn
   &\quad \quad \quad \quad + K_{\mathrm{bend}} 
   u_{\mathrm{bend}} \left(\bm{N}_{ij}^{(1)}, \bm{N}_{ij}^{(2)}\right) \bigg] \nn
   &+ \del\sum_{(i,j)\notin E_m}
   K_{\mathrm{ex}}u_{\mathrm{ex}}\left(\frac{|\bm{r}_i-\bm{r}_j|}{2R_m}\right),
   \label{eq:umm}
  \end{align}
  where $E_m$ represents the set of links of the basement membrane network. 
  
  The stretching energy is given by $u_{\mathrm{str}}(x)$:
  \begin{align}
   u_{\mathrm{str}}(x)=
   \begin{cases}
    \phi(\frac{x}{{\delta_2}}), & x < {\delta_2},\\
    \frac{\alpha_2}{2}(x-{\delta_2})^2 +\phi(1), & {\delta_2} \le x \le 1, \\
    \phi(\frac{1+{\delta_2}-x}{{\delta_2}}) + A_0+A_1 x, & 1 < x,
   \end{cases}
   \label{eq:stretch}
  \end{align}
  where the constants $A_0=-\frac{\alpha_2(1-{\delta_2}^2)}{2{\delta_2}^2}$ and $A_1=\frac{\alpha_2(1-{\delta_2})}{{\delta_2}^2}$ are chosen so that the function is $C^1$-continuous: See Fig.~\ref{fig:potential}(c). 
  
  The bending energy is also considered in each network link: Link $(i,j)$ uniquely determines a pair of two adjacent triangular regions, whose unit normals $\bm{N}_{ij}^{(1)}$ and $\bm{N}_{ij}^{(2)}$ in the case of Fig.~\ref{fig:model}(a) are given by $\bm{N}_{ij}^{(1)}=\frac{\bm{r}_i-\bm{r}_j}{|\bm{r}_i-\bm{r}_j|}\times\frac{\bm{r}_k-\bm{r}_j}{|\bm{r}_k-\bm{r}_j|}$ and $\bm{N}_{ij}^{(2)}=\frac{\bm{r}_j-\bm{r}_i}{|\bm{r}_j-\bm{r}_i|}\times\frac{\bm{r}_l-\bm{r}_i}{|\bm{r}_l-\bm{r}_i|}$. Here the sign of the normals is chosen so that they have positive $z$-components at the initial membrane
  configuration. The interaction function is defined as
  \begin{align}
   u_{\mathrm{bend}}(\bm{N}_{ij}^{(1)}, \bm{N}_{ij}^{(2)}) = \frac{1}{2}(1-\bm{N}_{ij}^{(1)}\cdot\bm{N}_{ij}^{(2)})^2.\label{eq:bend}
  \end{align}
  We choose $K_{\mathrm{str}}=0.04$, $K_{\mathrm{bend}}=0.5$, ${\delta_2}=0.25$, and $\alpha_2=1.25$. 
  
  \subsection{Cell-membrane adhesion} \label{sec:adhesion}
  The energy between the membrane and the cells are given as follows:
  \begin{align}
   U_{\mathrm{mc}} &= \sum_{i\in \Omega_{\TA}} K_{\mathrm{ad}}^{(i)} 
   u_{\mathrm{ad}} \left(\frac{|\bm{r}_i-\bm{r}_{\mathrm{nhd}(i)}|}{R_c+R_m}\right)\nonumber\\
   &+\sum_{i \in \Omega_{\stem}} K_{\mathrm{bind}}
   u_{\mathrm{bind}} \left(\frac{|\bm{r}_i-\bm{r}_{\mathrm{nhd}(i)}|}{R_c+R_m}\right)\nn
   &+ \sum_{i\in\Omega_c, j\in \Omega_m} K_{\mathrm{ex}} u_{\mathrm{ex}} \left(\frac{|\bm{r}_i-\bm{r}_j|}{R_c+R_m}\right),
   \label{eq:umc}
  \end{align}
  where $\Omega_{\TA}$ is the set of TA cells, $\Omega_{\stem}$ is the set of stem cells (including ND cells in Sec.~\ref{sec:nd}), and $\mathrm{nhd}(i) \in \Omega_m$ is the nearest membrane particle to $i$. The adhesion function depends on cell type, where
  \begin{gather}
   u_{\mathrm{ad}}(x)=
   \begin{cases}
    0, & x < 1, \\
    \frac{(x-1)^2}{2} - \frac{(x-1)^4}{4(\lambda-1)^2}, & 1\le x < \lambda, \\
    \frac{(\lambda-1)^2}{4}, & \lambda \le x.
   \end{cases} 
   \label{eq:ad} \\
   u_{\mathrm{bind}}(x)=
   \begin{cases}
    0, & x < 1, \\
    \frac{(x-1)^2}{2}, & 1\le x.
   \end{cases}
   \label{eq:ad_stem}
  \end{gather}

  We choose $K_{\mathrm{bind}}=25.0$, $K_{\mathrm{ad}}^{(i)}=25.0$ for proliferative TA cells and $K_{\mathrm{ad}}^{(i)}=0$ for non-proliferative TA cells. 
  
  \subsection{Cell division and cell-cell interactions}\label{sec:cell-cell}
  The total cell-cell interaction energy is given by
  \begin{align}
   U_{\mathrm{cc}} &= \sum_{\substack{i, j\in\Omega_c, i<j\\(i,j)\notin\Omega_{dp}}}
   \bigg[ 
   K_{\mathrm{ex}} u_{\mathrm{ex}}\left(\frac{|\bm{r}_i-\bm{r}_j|}{2R_c}\right) 
   + K_{\mathrm{cc}} u_{\mathrm{ad}}\left(\frac{|\bm{r}_i-\bm{r}_j|}{2R_c}\right) 
   \bigg]
   \nonumber\\
   &+\sum_{(i, j)\in \Omega_{dp}} K_{\mathrm{div}}\udiv\left(\frac{|\bm{r}_i-\bm{r}_j|}{2R_c}\right),
   \label{eq:ucc}
  \end{align}
  where $\Omega_c$ is the set of all cells, $\Omega_{dp}\subset \Omega_c \times \Omega_c$ is the set of dividing pairs (two cells committing the same division process). The same adhesion function \eqref{eq:ad} is used for cell-cell adhesion. For the interaction of a dividing pair, we adopt
  \begin{align}
   \udiv(x) = \frac{1}{2}\left(x - \beta_0(t-t_{ij})\right)^2,
  \end{align} 
  where $t_{ij}$ is the onset time of cell division between $i$ and $j$: A pair of cells are stored in $\Omega_{dp}$ when they start division and are removed when they complete division. 
  Note that the natural length of the spring between a dividing pair is given by $2R_c\beta_0(t-t_{ij})$. 
  
  We choose $\epsilon=0.005$, $\beta_0=0.14$, $K_{\mathrm{div}}=5.0$, $K_{\mathrm{ex}}=0.04$, and $K_{\mathrm{cc}}=0.03$. 
  
  \section{Cell cycle modeling}\label{sec:cellcycle}
  In this model, after passing a deterministic cell cycle with the period $T_0$, both TA and stem cells enter a stochastic division stage characterized by a Poisson process with the probability $\gamma$. The average division period therefore is given by $\av{T}=T_0+\gamma^{-1}$. 
  
  TA cell $i$ is assigned an integer $m_i$ representing a possible number of cell division ($0\le m_i \le M$), where we choose $M=10$. When cell $i$ with nonzero $m_i$ becomes $i$ and $j$ after division, both are assigned $m_i-1$; Cells created from stem cells are assigned $M$. Although it is known that stem cells have slower cell division cycle than TA cells, for the sake of simplicity we consider that stem cells and TA cells have the same cell division dynamics (same $\gamma$ and $T_0$). $T_0=36.0$ and $\gamma=0.083$ so that $\av{T}=48.0$.

  
 \end{document}